# Spin glasses : experimental signatures and salient outcomes

Eric Vincent [a] and Vincent Dupuis [b]

[a] SPEC, CEA, CNRS, Université Paris-Saclay, CEA Saclay, 91191 Gif-sur-Yvette Cedex, France
eric.vincent@cea.fr

[b] Sorbonne Universités UPMC Univ Paris 06 UMR 8234, PHENIX, F-75005 Paris, France
vincent.dupuis@upmc.fr

## Abstract

Within the wide class of disordered materials, spin glasses occupy a special place because of their conceptually simple definition of randomly interacting spins. Their modelling has triggered spectacular developments of out-of-equilibrium statistical physics, as well analytically as numerically, opening the way to a new vision of glasses in general. "Real" spin glasses are disordered magnetic materials which can be very diverse from the chemist's point of view, but all share a number of common properties, laying down the definition of generic spin glass behaviour. This paper aims at giving to non-specialist readers an idea of what spin glasses are from an experimentalist's point of view, describing as simply as possible their main features as they can be observed in the laboratory, referring to numerous detailed publications for more substantial discussions and for all theoretical developments, which are hardly sketched here. We strived to provide the readers who are interested in other glassy materials with some clues about the potential of spin glasses for improving their understanding of disordered matter. At least, arousing their curiosity for this fascinating subject will be considered a success.

# 1. Introduction

We are surrounded by disordered materials, in which the atoms or molecules are disposed at random. This is the case of window glass, but also of plastics, polymers, foams, gels, granular media, etc. Although being random when seen microscopically, they have controllable and reproducible properties at the macroscopic scale. Their modelling is a challenge for the material scientist as well as for the physicist.

Within the wide class of disordered materials, spin glasses appear as a remarkably simple archetype, because they can be defined in very simple terms. A spin glass is a set of interacting magnetic moments (originating from spins), in which the interactions are *randomly distributed in sign* (and possibly in magnitude). We easily represent ourselves ferromagnets (forming our permanent magnets), which are constituted of *positively* interacting moments, tending to all align in the same direction and produce a macroscopic magnetization. We also know antiferromagnets, in which the moments are in a *negative sign* interaction that drives them to anti-alignment, establishing a set of two intricated ferromagnetic sublattices oriented in opposite directions.

The case of spin glasses can be simply described as a mixture of both ferro- and antiferromagnetic situations. The magnetic moments (or spins) are in *random sign* interactions, that is, each moment experiences contradicting constraints from its neighbours, which are either ferromagnetically or antiferromagnetically interacting with it. This situation of contradicting influences has been termed *frustration*. No simple symmetric configuration of the set of spins corresponds to an equilibrium state with a clear minimum of energy. On the opposite, the numerous possible spin arrangements with comparable energy yield a huge number of metastable states. Finding the absolute minimum is thus extremely difficult and, from a practical point of view, a spin glass is virtually always out of equilibrium.

In a spin glass, the disorder is contained in the set of the magnetic interactions, which is fixed. Contrary to this situation of *frozen* disorder, in usual glasses the molecules are located at random positions that are evolving with time. The spin glass problem is conceptually simpler, it has allowed rich, far-reaching theoretical developments [1,2] and numerical investigations (see for instance the recent work [3] of the Janus collaboration, and references therein). Yet, both classes of systems share a lot of similitudes, and the spin glass has been progressively identified as a powerful model for the description and understanding of various glassy systems.

Disordered systems in which a cooperative behaviour is developing below a characteristic temperature are sometimes called "ferroic materials", a wide class of materials that constitute the subject of the book in which the present paper on spin glasses is a chapter. Interesting examples are martensitic alloys with shape memory effects [4], and relaxor ferroelectrics, on which some light can now be shed thanks to the analogy with certain spin glass models [5].

# 2. What is a spin glass made of?

The first spin glass materials identified were non-magnetic metals (Au, Ag, Pt…) in which a few percents of magnetic atoms (Fe, Mn…) were dispersed at random [6]. In Cu:Mn3 % for example, the Mn magnetic atoms are separated by random distances, and the oscillating character of their RKKY interaction with respect to distance makes their coupling constants take a random sign [7]. Examples of spin glasses have also been found within insulating compounds [8]. Interestingly, although chemically very different, these various compounds have been found to show a common general behaviour that is now understood as generic for spin glasses [9].

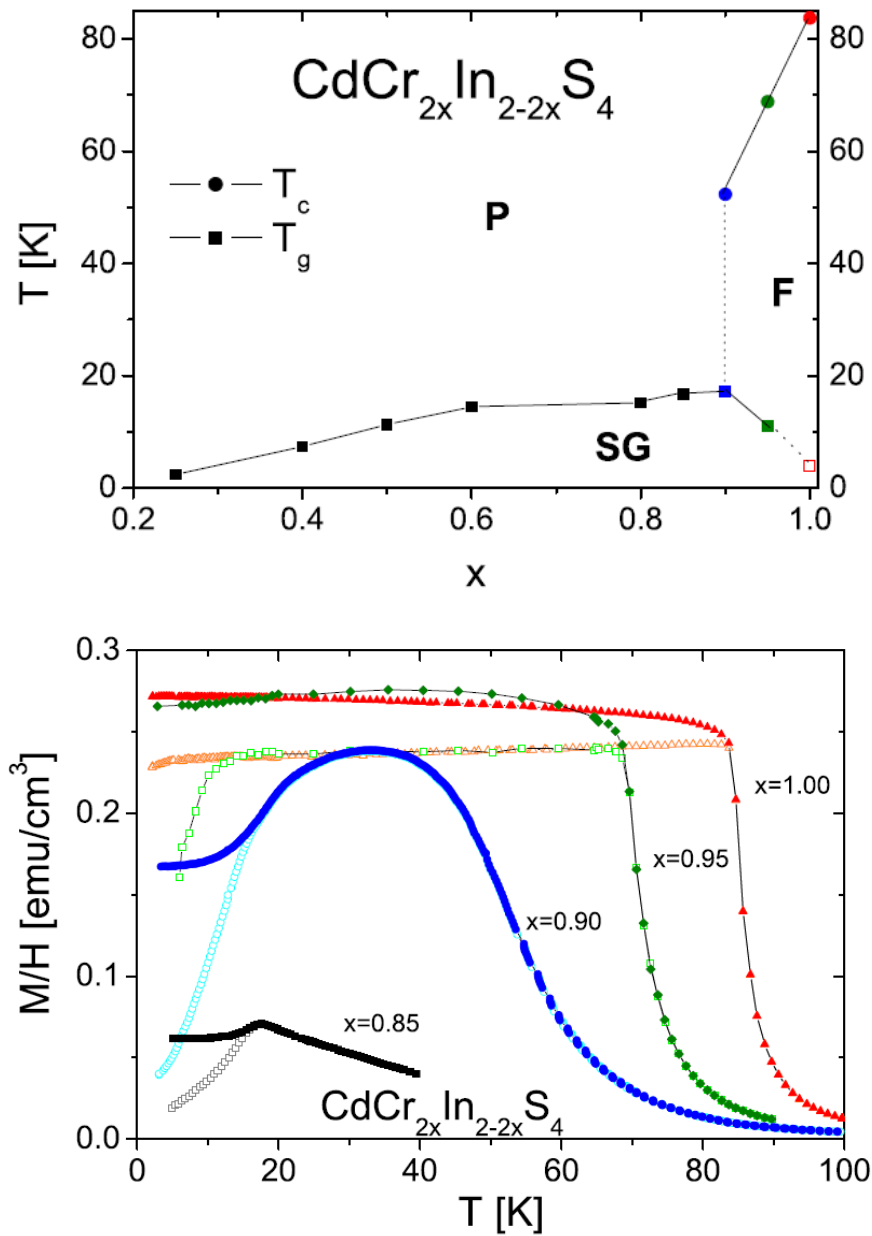

**Figure 1** : **Top:** Phase diagram of the $CdCr_{2x}In_{2(1-x)}S_4$ thiospinel compound, as a function of the dilution parameter x, showing the measured transition points between paramagnetic (P), ferromagnetic (F) and spin glass (SG) phases (lines are guides for the eye) [10,11]. **Bottom :** Magnetization (normalized to the field) as a function of temperature for five samples of the compound with various dilutions (the colours of the curves refer to the colours of the points in the phase diagram) [11]. The measurement follows the usual **ZFC** and **FC** procedures : for **ZFC**, cooling in zero field, applying the field at low temperature, then measuring upon increasing slowly the temperature, for **FC**, measuring upon slowly cooling in the field (the same curve is obtained upon re-heating).

An example in which a number of spin-glass properties have been observed in detail is the Indium diluted Chromium thiospinel CdCr$_{2x}$In$_{2(1-x)}$S$_4$, with superexchange magnetic interactions between the (magnetic) Cr$^{3+}$ ions [10]. The phase diagram is shown in Fig.1 (top) [11].

Let us first examine the x = 1 compound, which is a ferromagnet with $T_c$ = 85 K. The nearest neighbour interactions are ferromagnetic and dominant for x = 1, but the next-nearest ones are antiferromagnetic. Hence, there is some frustration even in the pure Cr compound. Characteristic variations of magnetization as a function of temperature are shown in Fig.1 (bottom), they are measured along the usual ZFC and FC procedures (see caption of Fig.1 for explanation). Starting from high temperatures, a rise-up of magnetization from the paramagnetic phase to the ferromagnetic plateau is clearly observed when approaching $T_c$ = 85 K. Below $T_c$, an irreversible behaviour is found, signed up by a separation of the FC and ZFC curves. The irreversibility, in which different geometrical arrangements of the ferromagnetic domains and walls are realized according to the temperature/field procedure, is probably due to some defects in the sample.

In CdCr$_{2x}$In$_{2(1-x)}$S$_4$, when a fraction (1-x) of the (magnetic) Cr ions is substituted by (non-magnetic) In ions, some nearest-neighbour *ferromagnetic* links are suppressed, and the effect of next-nearest *antiferromagnetic* interactions is enhanced [10]. The balance that globally favours ferromagnetism in the absence of In-dilution is disturbed. This is illustrated in Fig. 1 (bottom) in the case of the x=0.95 and x=0.90 samples, for which the ferromagnetic plateau becomes rounded. Meanwhile, the onset of irreversibility is shifted towards lower temperatures, indicating the appearance of a different, re-entrant spin glass phase at low temperatures (see phase diagram in top of Fig.1) [11].

For increasing dilution, below x ≤ 0.85, the ferromagnetic phase disappears. The transition occurs directly from the high-temperature paramagnetic phase to a disordered low-temperature phase, which presents all characteristic features of a spin glass, as will be explained below with various spin glass examples.

# 3. What happens at $T_g$?

## 3.a Dynamical aspects of the transition

When a structural glass is cooled down from its liquid phase, it fails to crystallize and becomes a supercooled liquid. The increase of relaxation times when cooling to the glass temperature is so abrupt that the supercooled liquid rapidly starts to behave as a good solid at all accessible experimental time scales [12,13,14]. For the so-called *fragile* glasses, which are the most common case, the viscosity (proportional to a typical relaxation time τ) of the supercooled liquid increases faster than the Arrhenius law corresponding to simple thermal slowing down over a barrier E :

$$\tau = \tau_0 \exp(E/k_B T),$$

$\tau_0$ being a microscopic time. This is pictured in the left part of Fig.2 [15], where the viscosity data from various glasses is presented. In this plot of the viscosity versus inverse temperature, the Arrhenius behaviour corresponds to a straight line, and most glasses show an upward curvature.

In a spin glass, there is also an abrupt increase of the relaxation times when cooling to the glass temperature. We know it quantitatively from the precise study of the magnetic ac susceptibility. At a fixed frequency $\omega/2\pi$, the ac susceptibility of a spin glass as a function of temperature presents a maximum at $T_f(\omega)$ that can even be very sharp [6]. $T_f(\omega)$ can be understood as the temperature at which the spin glass becomes frozen *at the experimental probing time scale $\tau$ equal to the inverse of the frequency, $\tau = 2\pi/\omega$*. This peak temperature is frequency dependent, and it systematically shifts to lower temperatures for decreasing frequencies, as can be seen in Fig.3 [16]. The important point is to quantitatively examine the scale of this frequency dependence, and to determine whether the peak temperature tends to a finite value in the limit of vanishing frequencies [17].

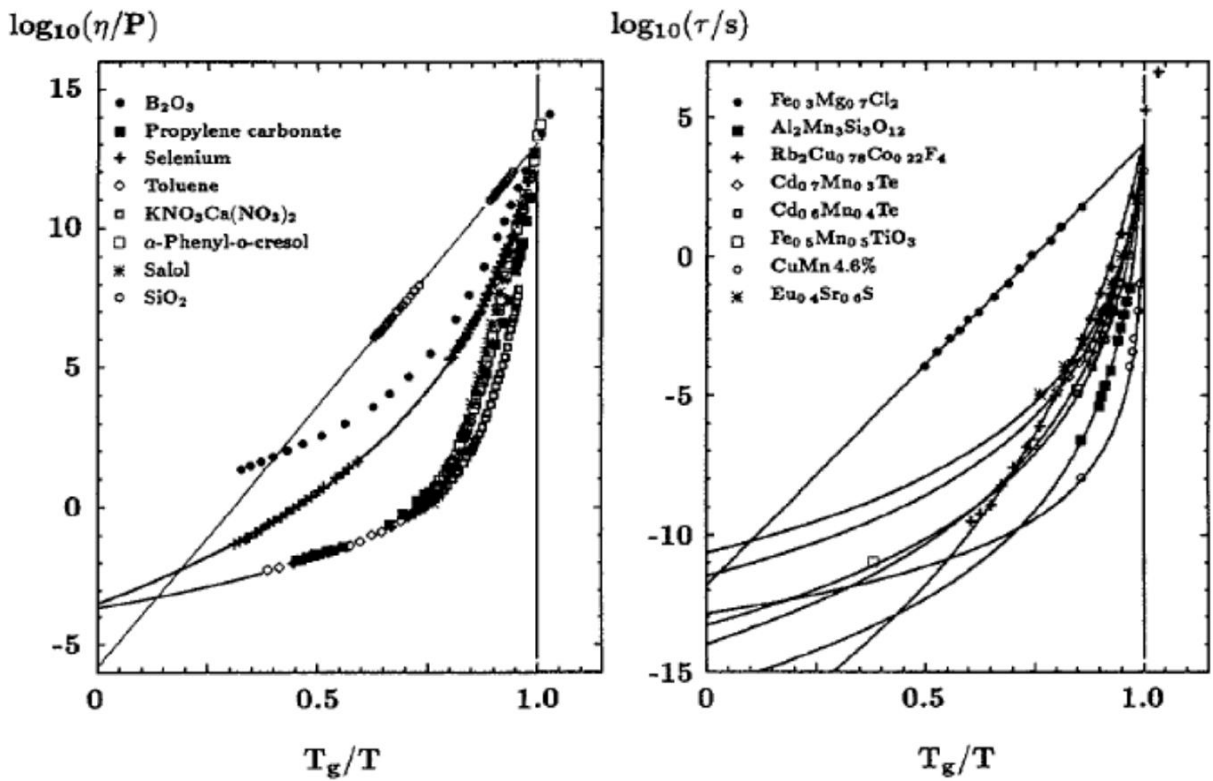

**Figure 2** : (from [15]) **Left** : Plot of the logarithm of the viscosity $\eta$ of different glass-forming systems vs. $T_g/T$, the glass temperature $T_g$ being defined such that $\eta = 10^{13}$ Poise at $T_g$. Only $SiO_2$ follows an Arrhenius behaviour (straight line in this plot, "strong" glass). The more common "fragile" glasses show a viscosity increase towards low temperatures that is faster than Arrhenius. **Right** : Plot of the logarithm of a typical relaxation time $\tau$ of different spin glasses vs. $T_g/T$, the glass temperature $T_g$ being defined such that $\tau = 10^4$ s at $T_g$. Most standard spin glasses appear as "fragile", according to the classification of glasses. In more details, spin glasses usually obey critical dynamic scaling (see text), and for structural glasses the question of criticality implies further investigations [18].

The shift of $T_f(\omega)$ with $\omega$ in spin glasses can be regarded as an increase of the value of a typical relaxation time $2\pi/\omega$ for decreasing temperature $T_f(\omega)$. It can be presented in the same kind of Arrhenius plot of time versus inverse temperature as is used for glasses. This is shown in the right part of Fig.2. Both Arrhenius plots for glasses and spin glasses look very similar. In both cases the increase of the relaxation times with decreasing temperature is faster than an Arrhenius law (upward curvature) [15].

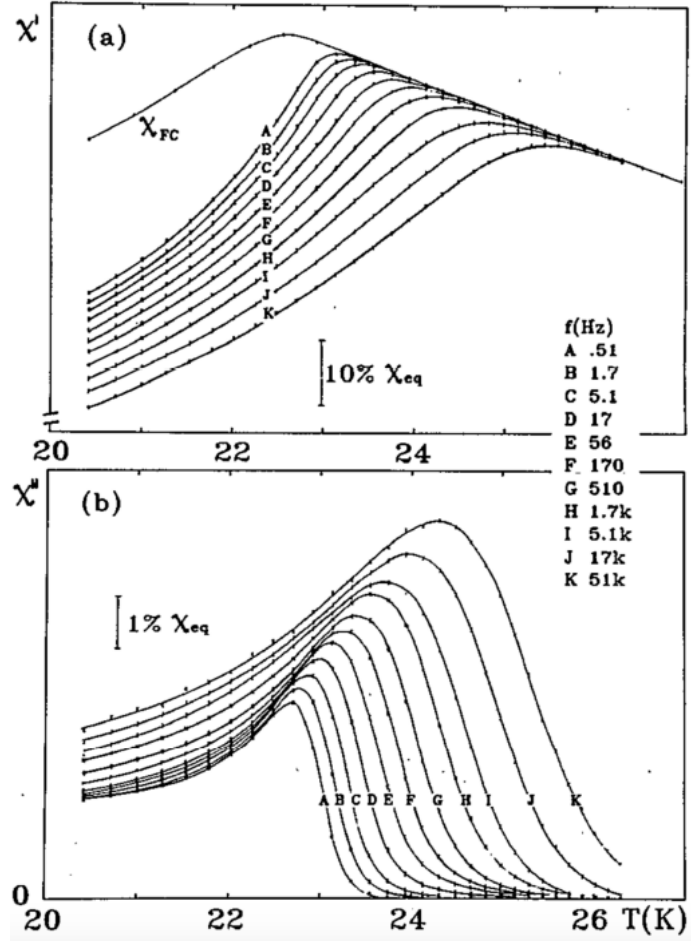

**Figure 3** : (from [16]) Complex *ac* susceptibility $\chi(\omega) = \chi'(\omega) + i\chi''(\omega)$ of the amorphous metallic spin glass $(Fe_xNi_{1-x})B_{16}P_6Al_3$, versus temperature, at different frequencies (0.51 Hz to 51 kHz) of the applied oscillating field (amplitude 50 mOe). **(a)** $\chi'(\omega)$. The field-cooled susceptibility $\chi_{FC}$ at an applied field of 50 mOe is also plotted. 10% of the equilibrium susceptibility ($\approx \chi_{FC}$) is indicated. **(b)** $\chi''(\omega)$. 1% of the equilibrium susceptibility ($\approx \chi_{FC}$) is indicated. The frequency dependent freezing temperature $T_f(\omega)$ can be defined equivalently as the temperature of the $\chi'(\omega)$ peak or as that of the $\chi''(\omega)$ inflection point.

Numerically, the curves in Fig.2 can be well fitted to the Vogel-Fulcher law [19]

$$\tau = \tau_0 \exp(E/k_B(T-T_0)) ,$$

where $T_0$ is an adjustable parameter. But, for spin glasses, nothing particular is happening at $T_0$, to which no physical interpretation is usually given (the situation is different for structural glasses [20]). It is instructive to simply consider that the departure of the data from the Arrhenius law corresponds to a modification of the Arrhenius law by the introduction of a *temperature dependent* effective energy barrier $E_{eff}(T)$ :

$$\tau = \tau_0 \exp(E_{eff}(T)/k_BT) .$$

In these terms, the upward curvature of the data in Fig.2 means an increase of $E_{eff}(T)$ for decreasing temperatures, which can be considered as a signature of the development of correlations when approaching $T_g$ from above, both in glasses and in spin glasses.

In glass-forming liquids such as glycerol, an increase of the number of correlated molecules has now been identified at the approach of $T_g$ [18]. This increase remains limited to a relatively modest extent of a few tens of molecules before dynamical arrest, but it paves the way to a new understanding of the glass transition in terms of an increase, even though limited before freezing, of a correlation length at the approach of the glass transition [21].

In spin glasses, the increase of a relaxation time $\tau = 2\pi/\omega$ for decreasing temperatures (data in Fig.2 right, obtained from measurements like those presented in Fig.3) can be well fitted considering a divergence of a correlation length $\xi$ at the approach of a transition at $T_g$,

$$\xi = \xi_0 [ (T_f(\omega)-T_g)/T_g ]^{-\nu}$$

($\nu$ being the usual exponent for the correlation length in a phase transition), and using the dynamic scaling hypothesis

$$\tau \propto \xi^z$$

($z$ is thus defined as a dynamical exponent), which yields the critical dynamics scaling law :

$$\tau = \tau_0 [ (T_f(\omega)-T_g)/T_g ]^{-z\nu} \quad [22].$$

The exponent $z\nu$ is found to have a rather high value, ranging from 5 to 11 in the various samples (see for example [23,16]).

## 3.b A thermodynamic phase transition

There are other classes of experiments in spin glasses which support the idea of a thermodynamic phase transition at the zero-frequency limit $T_g$ of the freezing temperature $T_f(\omega)$. In a ferromagnet, the order parameter is the spontaneous magnetization, and the order parameter susceptibility is the usual magnetic susceptibility. In a spin glass, some "glassy order" takes place, yielding to an apparently random arrangement of the spins with no visible macroscopic symmetry. The low temperature phase can be characterized by the Edwards-Anderson order parameter [24], which corresponds to an average of the squared moduli of the spins, and the order parameter susceptibility is the non-linear magnetic susceptibility [25].

The magnetic susceptibility $\chi$ can be expanded in even powers of the magnetic field $H$ :

$$\chi = \chi_0 - a_3 H^2 + a_5 H^4 ... ,$$

$\chi_0$ being the linear susceptibility. The coefficients of the non-linear terms are all diverging at $T_g$, with critical exponents corresponding to the specific spin glass order parameter [25]:

$$a_3 \propto (T-T_g)^{-\gamma}, a_5 \propto (T-T_g)^{-(\beta+\gamma)}, \text{etc.}$$

Their determination implies rather extensive measurements of the magnetic susceptibility as a function of the field, at various temperatures close to $T_g$, and careful extrapolation at zero field. Figure 4 shows the example of Ag:Mn$_{0.5\%}$ [26]. The non-linear part of the susceptibility is plotted as a function of the field. In this figure, a significant increase of the slope of the curves at the origin for decreasing values of the temperature $T$ towards $T_g$ is very clearly visible. Such evidences of a static critical behaviour from the non-linear susceptibility, in addition to dynamic critical behaviour determined from the *ac* susceptibility, have been obtained in numerous different spin glass samples (see for example [26,27,28,29], and numerous references in [9]).

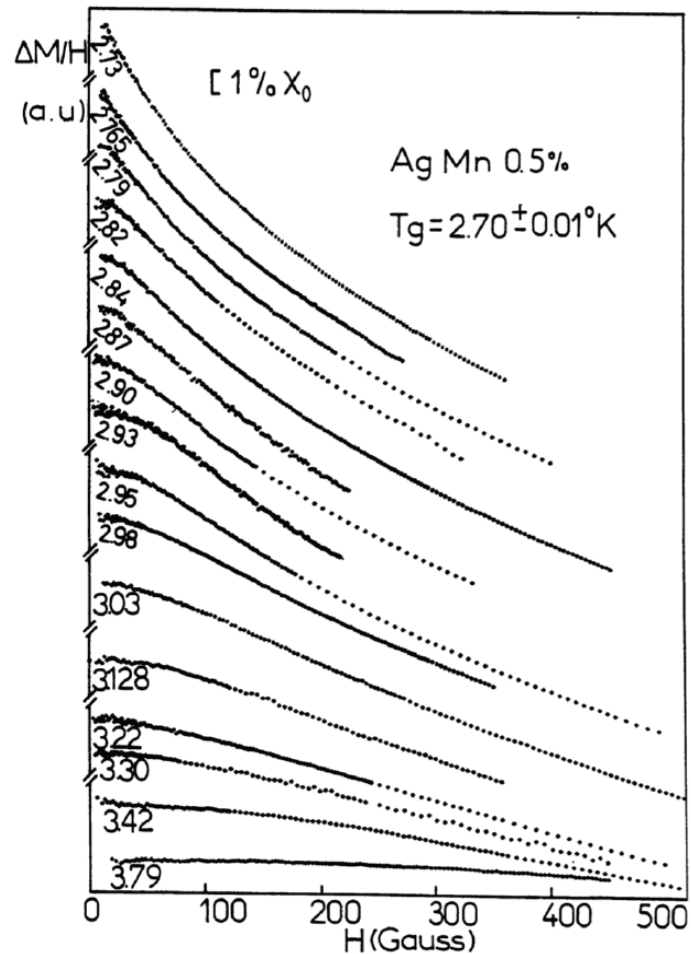

**Figure 4 :** (from [26]) Non-linear susceptibility (obtained as the difference $\Delta M$ between the total magnetization and its field-linear part, divided by the applied magnetic field *H*), as a function of field *H*, at different temperatures approaching the critical temperature $T_g$ =2.70K. The relative origins on the Y-axis are arbitrary. 1% of the linear susceptibility $\chi_0$ is indicated. As $T \rightarrow T_g$, a sharp increase of the slope at the origin is clearly visible. Below $1.1T_g$=2.97K, the low field behaviour of the non-linear susceptibility is seen to become singular instead of being quadratic.

### 3.c Spin glass transition : open questions

The spin glass transition in real samples is now widely considered as a thermodynamic phase transition, in agreement with mean-field spin glass models [30,2]. Still, a few interesting questions are worth being mentioned on the subject. In the mean-field theory [2], which is equivalent to infinite dimension *d*, a true phase transition is indeed obtained, which persists in the presence of a magnetic field, as well for scalar Ising [31] as for vector Heisenberg [32] spins. In *d*=3, a phase transition is expected for Ising spins, but not for Heisenberg spins. However, many evidences of a phase transition are found in real *d*=3 Heisenberg-like samples (e.g. [27], see references in [9]). A very plausible explanation can be found in the scenario proposed by Kawamura of chirality driven phase transition of spins, a mechanism that has been detailed and argued both numerically and analytically [33]. The agreement between the values of the critical exponents found in this scenario and those observed in the experiments is remarkable [29]. Experimentally, direct access to the observation of chirality freezing is difficult, but some pioneering measurements of the anomalous Hall effect in spin glasses have already given very interesting results [34,35].

At the spin glass transition, critical exponents which vary from Ising to Heisenberg situations have been measured in a wide series of *d*=3 samples with variable spin anisotropy [29,9]. Experiments on spin glass *thin films* have allowed to study the crossover from *d*=3 to *d*=2, a situation in which the transition is expected to take place at T=0, and which allows interesting studies of the growth of the spin-glass correlation length under constrained conditions (see recent results in [36], and older references therein).

In the mean-field theory of spin glasses [2], an *infinite number of different pure states* is obtained, yielding a very interesting and complex picture of the spin glass phase. This would imply that in a real sample, after cooling from the paramagnetic phase, many domains with different types of spin glass order should coexist and compete. A phase diagram with a transition line is obtained as a function of the magnetic field [31,32]. On the other hand, in very different approaches, scaling theories of the spin glass behaviour have been developed for Ising spins on the basis of phenomenological arguments [37,38]. In the droplet model [37], there are simply two (spin reversal symmetric) pure states, and *the phase transition is expected to be destroyed by any magnetic field*. Let us comment briefly on these important questions.

(i) The question of a multiple or single nature of the ground state in the spin glass is very difficult to address experimentally. Some (indirect) arguments in favour of multiple pure states are discussed below (Section 5c). Also, following a theoretical suggestion stating that the correlation of the conductance fluctuations in two spin states should be a direct function of their overlap [39], a new experimental approach using transport measurements on mesoscopic samples has been developed. Magneto-resistance traces are observed, which are reproducible until the sample is heated well over $T_g$. They are likely to be correlated to frozen spin configurations which strikingly persist under high field cycling [40]. Measurements of the universal conductance fluctuations in mesoscopic spin glasses is a real challenge, and have not yet given full results, but in principle they are a promising way to obtain information on the nature of the pure states.

(ii) The vanishing of the phase transition in presence of a magnetic field has been reported in a study of the dynamic scaling properties of a $Fe_{0.5}Mn_{0.5}TiO_3$ single crystal [41], which is a good example of a short-range Ising spin glass [42]. Interestingly, very recent experiments on the $Dy_xY_{1-x}Ru_2Si_2$ show that the phase transition persists in a field in this Ising *but long-range* (RKKY) system [43]. For Heisenberg-like spin glasses, data from torque measurements bring robust evidence for a true spin glass ordered state which survives under high applied magnetic fields [29].

Thereby, important questions concerning the nature of the spin glass transition are still open. They are also a hot topic for structural glasses, in which the non-linear susceptibility is now understood as playing a similar role as in spin glasses [21]. An important experimental program has now allowed investigating, by the means of non-linear dielectric susceptibility measurements, the growth of correlations at the approach of the transition and in the glassy phase during aging [18].

In experiments on spin glasses, no true thermodynamic equilibrium state can be reached at laboratory time scales. What we see in experiments probing the spin glass state is essentially out-of-equilibrium properties [44]. A wide panel of rich results have been obtained, of which we highlight in Sections 4 and 5 some of the most prominent features.

# 4. Slow dynamics and aging

## 4.a *DC* experimental procedures

We present in Figure 5 a typical measurement of the magnetization as a function of temperature in a spin glass [45], performed along the usual ZFC-FC procedures (see caption of Fig.1). The sample is here a $Fe_{0.5}Mn_{0.5}TiO_3$ single crystal, which is a good example of a spin glass of Ising type, due to strong easy axis anisotropy [42].

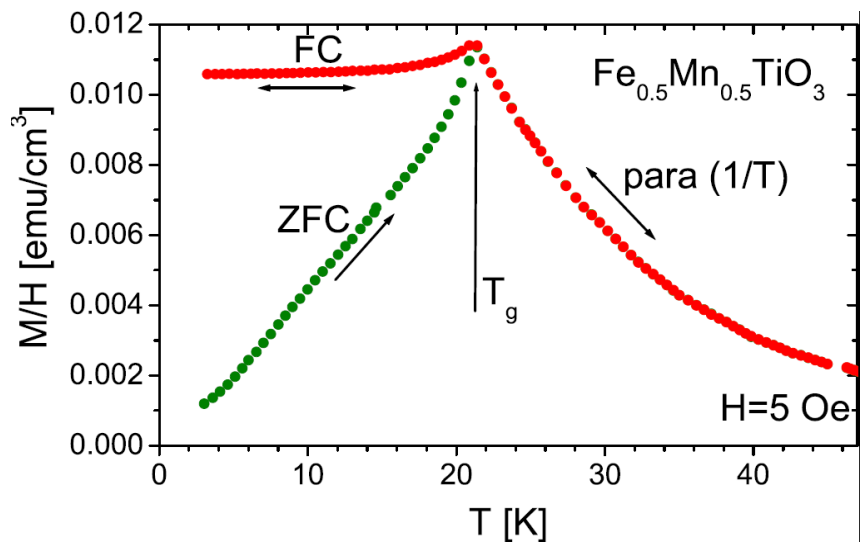

**Figure 5 :** (from [45]) Magnetization divided by the applied field, as a function of temperature, measured along the usual ZFC-FC procedures (see caption of Fig.1) on a a $Fe_{0.5}Mn_{0.5}TiO_3$ single crystal [42] along the *c*-axis direction. The onset of irreversibility is seen at $T_g$ in the separation of the ZFC-FC curves, which above $T_g$ are identical with a 1/T-like paramagnetic behaviour.

The curves are very similar to those presented in Fig.1 for the x=0.85 diluted thiospinel sample, they illustrate the general features of simple *dc* magnetization measurements on spin glasses. Above $T_g$, the magnetization *M* follows a characteristic Curie-Weiss law

$$M \propto C/(T-\theta)$$

which is characteristic of a paramagnetic phase (*C* is a constant proportional to the square of the individual magnetic moments, and $\theta$ is a temperature proportional to the energy of the interactions). Below $T_g$, the temperature behaviour becomes irreversible. The magnetization is history-dependent, and a splitting of the ZFC and FC curves is observed. Such a splitting can be found in various magnetic systems in which some freezing occurs for any reason (e.g. non-interacting magnetic nanoparticles, see a review in [46]). What is characteristic here of the collective behaviour related to the spin glass transition is the (approximate) flatness of the FC curve observed here below $T_g$ (also obtained as a characteristic feature in mean-field models [2], as emphasized in [47]). When going from the paramagnetic region to low temperatures, the magnetization increase suddenly stops.

In the FC state, the magnetization value can be considered to a first approximation to be at equilibrium (this is usually true within 1%), and the FC curve can be measured upon cooling or as well heating in presence of the field. On the contrary, the ZFC curve is out of equilibrium, because the application of the field has been made at low temperature, in the frozen phase. The value of the ZFC magnetization depends on the time spent at each temperature. After cooling in zero field and applying the field, the ZFC($t$) magnetization slowly increases as a function of time, most probably towards the FC value (that is, however, never attained in experimental time scales).

An example is shown in Fig. 6 [16], in which we can see that the relaxation curves are influenced by another time parameter: the waiting time.

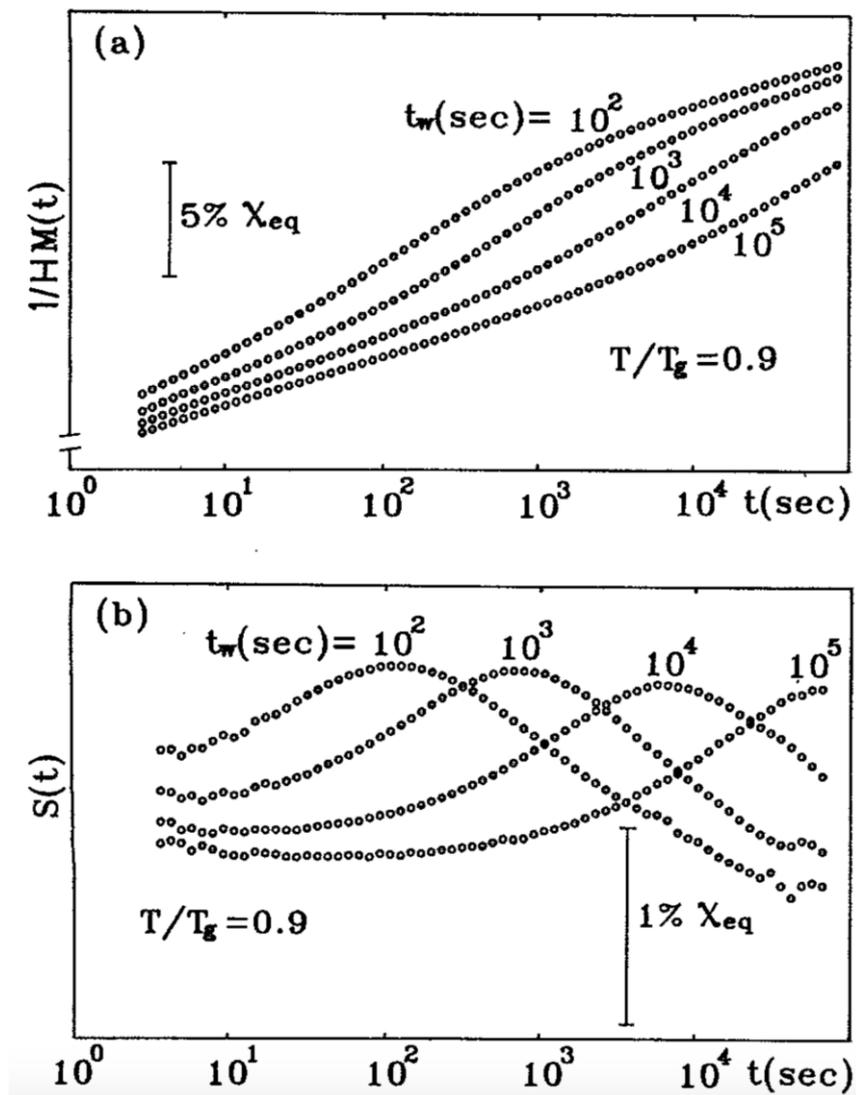

**Figure 6 :** (from [16]) Zero-field cooled susceptibility [$(1/H)M(t)$] and corresponding relaxation rate [$S(t)=(1/H)dM/d\ln t$] at different waiting times ($t_w$ =$10^2$, $10^3$, $10^4$, and $10^5$ s) plotted versus time $t$ (log scale), for the amorphous metallic spin glass (Fe$_x$Ni$_{1-x}$)B$_{16}$P$_6$Al$_3$ at $T$=20.3 K ($T/T_g$=0.9), $H$=0.1 Oe. **Top :** [$(1/H)M(t)$], and **Bottom :** $S(t)$. 1% of the equilibrium susceptibility ($\approx \chi_{FC}$, see Fig.3 on the same sample) is indicated.

In the procedure used to measure these relaxations (Fig.6), the sample is rapidly cooled in zero field from above $T_g$ to $T<T_g$ (quench), and it is kept at temperature T during a given *waiting time* $t_w$, after which a small field is applied (defining *t=0*). Then the relaxation is measured as a function of the observation time *t* (elapsed since the field change).

The relaxation curves in Fig. 6 (top) display two essential features of spin-glass dynamics:

(i) the magnetization relaxation following a field change is slow, roughly logarithmic in time (glassy state),

(ii) it strongly depends on the waiting time: the longer $t_w$, the slower the relaxation. This waiting time dependence is called *aging*.

Hence, time translation invariance is lost in the slow dynamics of the spin glass: the relaxation dynamics depend on *both $t_w$ and t*, not only on *t*, the dynamics is non-stationary [48].

The logarithmic derivative *dM/d log t* of these magnetization curves can be given an interesting physical interpretation, as proposed by L. Lundgren at the beginning of the 80's [49]. The derivatives of the curves of the top part of Fig. 6 are shown in the bottom part. They are bell shaped, and, remarkably, their broad maximum (inflection point of the magnetization curves) occurs after a time *t* of the order of $t_w$ itself.

The physical interpretation is the following. As can be seen on the *log t* scale, the relaxations are slower than exponential. They can be modelled as a sum of exponential decays *exp(-t/τ)*, the decay times *τ* being distributed according to a distribution function *g(τ)* which is defined in this way, and represents the effective density of relaxation times. Taking the derivative *dM/d log t* introduces a term *t/τ exp(-t/τ)* in the integrand of the sum over the *τ* distribution. This term is sharply peaked around *t=τ*. Approximating this peaked function by a δ-function, we estimate the integral by taking the value of the integrand for *τ=t*, and obtain [49]

$$dM_{tw}/d\ log\ t\ \propto\ g_{tw}(\tau=t)\ .$$

Here $M_{tw}$ and $g_{tw}$ are labelled by $t_w$ to emphasize that each relaxation curve, taken for a given $t_w$, gives access (through its logarithmic time derivative) to the density of relaxation times that represents the dynamics of the spin glass at a time of the order of $t_w$ after the quench. Thus, each derivative *dM$_{tw}$/d log t* gives an estimate of the density $g_{tw}(\tau=t)$, and as $t_w$ increases $g_{tw}(\tau)$ shifts towards longer times. This gives a physical picture of the two important features listed above:

(i) the effective relaxation times are widely distributed (glassy state)

(ii) this distribution function, peaking around *τ=$t_w$*, shifts towards longer times with increasing $t_w$ : this is the phenomenon of aging.

The mirror experiment of the above ZFC relaxations can be performed as well, it gives access to the same information, provided that the amplitude of the field change remains small (far below the limit of linearity, typically 1-10 Oe). In this mirror procedure, one starts from a FC state at temperature *T>$T_g$*. After cooling the spin glass from the paramagnetic phase to a temperature *T< $T_g$*, and waiting a time $t_w$ at *T,* the field is turned to zero. Then the remnant magnetization (called "thermo-remnant" magnetization, TRM) slowly decays [50].

Both relaxations observed by ZFC and TRM mirror procedures are symmetric: ZFC($t$) + TRM($t$) = FC (this relation holds even if a slight relaxation of the FC magnetization occurs, FC ≡ FC($t$) [51]). All these curves present an inflection point at $t\sim t_w$. When plotted as a function of $t/t_w$, the curves can be almost superimposed. In a first approximation, we can thus consider that the relaxation curves obey a $t/t_w$ scaling. When examined in more details, however, some systematic departures from $t/t_w$ scaling are observed, and can be taken into account very precisely by more refined procedures ("subaging", [50,52,53]).

The same phenomenon of aging has been known for a long time in the mechanical properties of a wide class of materials called "glassy polymers" [54,55]. When a piece of e.g. PVC is submitted to a mechanical stress, its response (elongation, torsion …) is logarithmically slow. And the response depends on the time elapsed since the polymer has been quenched below its freezing temperature. Like in spin glasses, for increasing aging time the response becomes slower (called "physical aging", as opposed to "chemical aging").  The $t_w$-dependence of the dynamics of glassy polymers has been expressed as a scaling law [54] that could be applied to the case of spin glasses ([50], see also [56]). Numerous other glassy materials show similar aging phenomena, although not necessarily obeying precisely a $t/t_w$ scaling (see for example [54,55,57,58]). Numerical simulations of packed hard spheres provide us with very powerful toy models of simple glasses [59].

## 4.b *AC* experimental procedures

Slow dynamics and aging in the spin-glass phase can also be observed by *ac* susceptibility measurements, in which a small *ac* field (~1 Oe) is applied all along the measurement [50,44]. Again, the starting point consists in cooling the spin glass from above $T_g$, down to a given $T<T_g$ at which the *ac* response is measured as a function of the time elapsing, which is the "age" of the system (equivalent to $t_w+t$ in the *dc* procedures).

Figure 7 [45] shows the time evolution of both components $\chi'$ and $\chi''$ of the ac susceptibility. We find here the same features as observed in *DC* experiments:

(i) The *ac* response is delayed, as seen from the existence of an out-of-phase susceptibility $\chi''$. $\chi''$ is zero above $T_g$ in the paramagnetic phase, and rises up as the sample is cooled into the spin-glass phase (as already visible in Fig. 3).

(ii) The susceptibility relaxes down, signing up the occurrence of aging. This relaxation is visible on both $\chi'$ and $\chi''$, but is more important in relative value $\Delta\chi/\chi$ in the out-of-phase component $\chi''$.

In Figure 7, the relaxations of $\chi'$ and $\chi''$ are shown for different frequencies $\omega/2\pi$, and plotted as a function of $(\omega/2\pi).t$ for reasons that will soon become clear. Their asymptotic (stationary) values $\chi'_{eq}(\omega)$ and $\chi''_{eq}(\omega)$ in the infinite time limit can be determined by a fit of the decaying part to $(\omega.t)^{-b}$ (the exponent $b$ is found in the range 0.15-0.20 in various samples [45]).

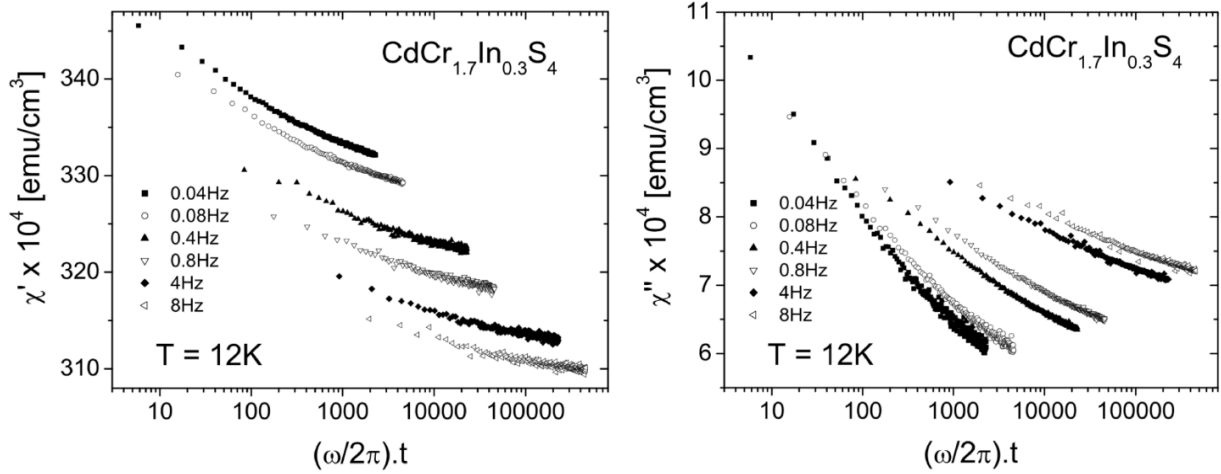

**Figure 7** : (from [45]) Relaxation of both components of the *ac* susceptibility during the time *t* following the quench from above $T_g$ down to $T=12K=0.7T_g$, as a function of the product of the frequency $\omega/2\pi$ times *t*. **Left :** in-phase component $\chi'$. **Right :** out-of-phase component $\chi''$.

In Figure 8 [45], the asymptotic values have been subtracted, and the remaining decaying part is represented on a log scale, which emphasizes the power law behavior (straight lines in this log-log plot). Remarkably, the decay part of the curves measured at different frequencies are all superimposed when plotted as a function of $(\omega/2\pi).t$, as well for $\chi'$ as for $\chi''$.

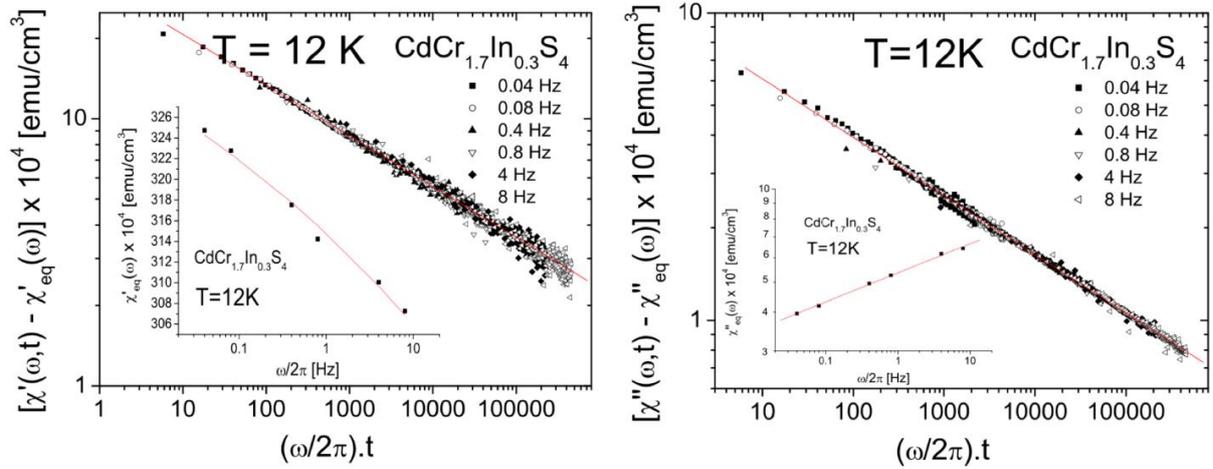

**Figure 8** : (from [45]) Relaxation of both components of the *ac* susceptibility, same data as in Fig.7, but after subtraction of the equilibrium part, log-log scale. The inserts show the fitted equilibrium values. The relaxations at different frequencies $\omega/2\pi$ merge onto a unique curve (power law) as a function of the reduced variable $(\omega/2\pi).t$.

This $\omega.t$ scaling can be related to the $t/t_w$ scaling of the *dc* data in the following way. In an *ac* measurement, the time $2\pi/\omega$ can be considered as a typical observation time, which plays the same role as *t* in the *DC* relaxation procedures. On the other hand, the total age of the system is here the time *t* along which the *ac* relaxation is measured after cooling, that is equivalent to $t_w+t$ in the *DC* experiment. Hence,

$$\omega.t \approx (1/t)(t+t_w) = 1 + t_w/t,$$

the present $\omega.t$ scaling is equivalent to the $t/t_w$ scaling of the *dc* experiments [49,50,52,44].

In structural and polymer glasses, similar *ac* procedures have also been used for the study of aging effects. For instance, in [60], the dielectric constant ε of glycerol is found to show a strong relaxation, with a frequency dependence which has qualitatively the same trend as $\omega.t$ scaling. And, more recently, detailed measurements of the third harmonics dielectric susceptibility in glycerol [61] have revealed an increase of the size of the glassy domains with the aging time.

Beyond the study of the response of the spin glass to small *dc* and *ac* excitations, the dynamics can also be explored by measuring the spontaneous magnetic fluctuations (magnetic noise) in the absence of any excitation. This is a difficult experiment, because contrary to the case of the ferromagnet the amplitude of the spontaneous noise is very small in spin glasses. Nevertheless, such measurements could be performed [48,50,62]. They brought very interesting information on the violation of the fluctuation-dissipation relations in the aging regime, which allow comparisons with some important features of spin-glass theories ([63], see also Figs. 5-7 in [47]).

# 5. Aging, rejuvenation and memory effects

It is well known that the state of a structural glass is very much dependent on the speed at which it has been cooled, a slower cooling bringing smaller values of the enthalpy and specific volume that are closer to equilibrium values [12,64 ,54]. This view of glasses was the starting point of a class of experiments in spin glasses. We explored how the aging behaviour could be influenced by the temperature history, having in mind that well-suited cooling procedures might perhaps bring the spin glass into a strongly aged state, which otherwise would require astronomical waiting times to be established [65,66]. These experiments have brought important surprises [44].

## 5.a Temperature step experiments

Figure 9 presents the result of an experiment in which a small negative temperature cycle is performed during the relaxation of the *ac* susceptibility [66]. After a normal cooling (~100 s from *1.3 $T_g$* to *0.7 $T_g$*), the spin glass is kept at constant temperature *T = 12 K = 0.7 $T_g$* for *$t_1$ = 300* minutes, during which aging is visible in the strong relaxation of $\chi''$. Then, the temperature is lowered one step further from *T = 12 K* to *T-ΔT = 10 K*. What is then observed is not a slowing down of the relaxation, but on the contrary a jump of $\chi''$ and a restart. Such a restart upon further cooling was termed *rejuvenation*, because the relaxation of $\chi''$ behaves as if aging was starting anew at *T-ΔT*. Apparently, there is no influence of former aging at *T*.

The question one may naturally ask is whether this renewed relaxation corresponds to a *full rejuvenation* of the sample. The answer is no. Let us first point out that, for observing such a restart, the temperature interval *ΔT* must obviously be sufficiently large, here *ΔT ≥ 2-3 K*. And still, the time window explored in this experiment is limited, therefore we do not know very much about the *overall* state of the spin glass, which involves relaxation processes on a very wide time scale.

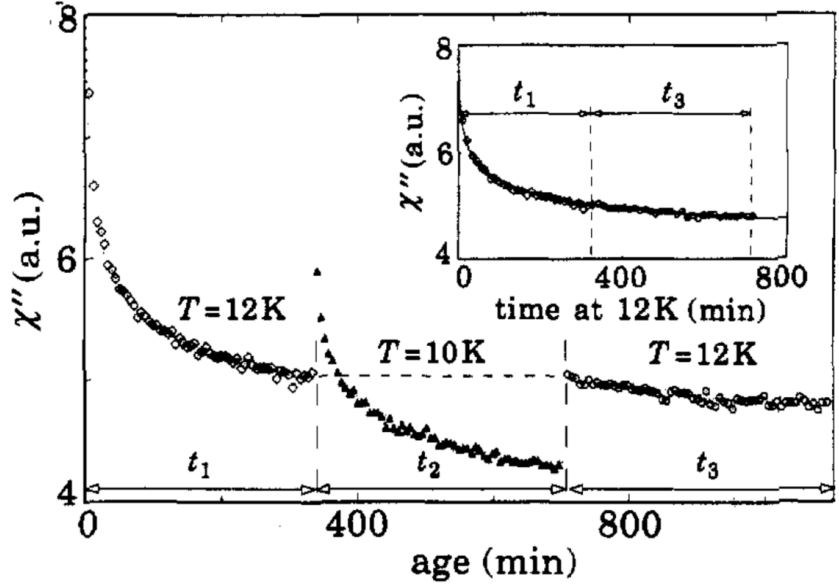

**Figure 9 :** (from [66]) Relaxation of the out-of-phase susceptibility $\chi''$ during a negative temperature cycle of amplitude $\Delta T$=2K (frequency 0.01 Hz), showing aging at 12 K, rejuvenation at 10 K, and memory at 12 K. The inset shows that, despite the *rejuvenation* at 10 K, both parts at 12K are in continuation of each other (*memory*). The sample is the CdCr$_{2x}$In$_{2(1-x)}$S$_4$ thiospinel spin glass ($T_g$=16.7K).

The final part of the experiment brings the answer to the question. After aging during $t_2 = 300$ min at $T$-$\Delta T = 10\ K$, when the temperature is turned back to $T = 12\ K$, the $\chi''$ relaxation restarts exactly from the point that was attained at the end of the stay at the original temperature $T$. It goes in precise continuity of the former one, as if nothing of relevance at $T$ had happened at $T$-$\Delta T$. As shown in the inset of Fig.9, this can be checked by shifting the 3$^{rd}$ relaxation to the end of the 1$^{st}$ one: they are in continuity, and can be superposed on the reference curve which is obtained in a simple aging at $T$. Hence, during aging at $T$-$\Delta T$ and despite the strong associated $\chi''$-relaxation, the spin glass has kept a *memory* of previous aging at $T$. This memory is retrieved when heating back to $T$.

This negative temperature cycle experiment pictures in a spectacular manner the phenomenon of rejuvenation and memory in a spin glass. However, examination of the situation in more details shows that it should not be considered too simply. We see in Figure 10 the results of negative temperature cycle experiments performed with various values of $\Delta T$ ([45], but see also [67]).

For $\Delta T = 1\ K$, the beginning of the 3$^{rd}$ part relaxation shows a transient spike, which lasts for ~ 5000 s before the curve merges with those, obtained for higher $\Delta T$'s, that are in continuity with the relaxation at $T$. Thus, for a smaller $\Delta T$ than that corresponding to a "full" memory effect, there is indeed some contribution at $T$ from aging at $T$-$\Delta T$, that may be divided in two parts:

1. an *incoherent* contribution (spike), extending over rather long but finite times (3-5000 s). For smaller $\Delta T$, the observed "transient spike" decreases, and finally vanishes,
2. a coherent contribution to aging at $T$ as an additional aging time $t_{eff}$ (called "cumulative process" below), in such a way that the 3$^{rd}$ relaxation must be shifted by $(t_2$-$t_{eff})$ to be in continuity with the 1$^{st}$ part.

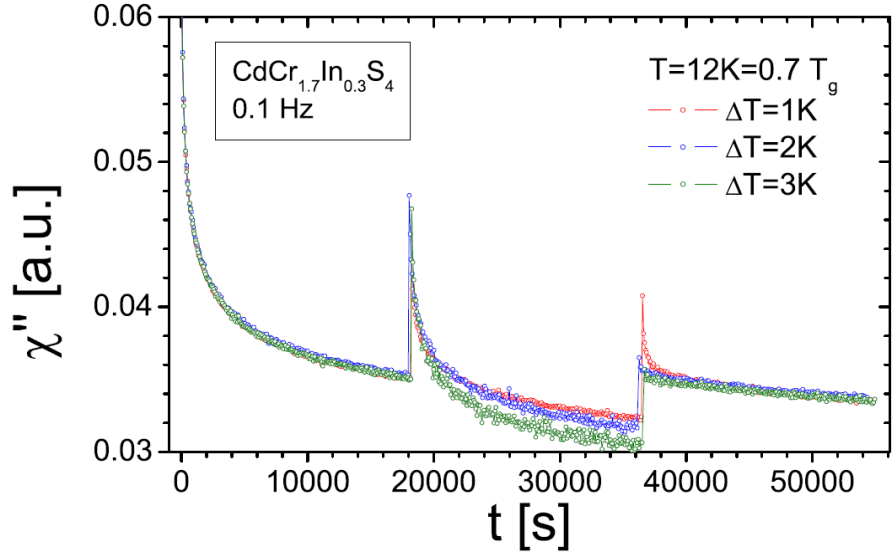

**Figure 10 :** (data from [45]) Relaxation of the out-of-phase susceptibility $\chi''$ during negative temperature cycles of different amplitudes, ranging from $\Delta T$=1K (upper curve, with the prominent spike) to $\Delta T$=3K (lower curve, no spike and full memory). Same sample as in Fig.9, but the frequency is here 0.1 Hz, instead of 0.01 Hz (see note [68] for comparison).

Details on the results in this regime of intermediate $\Delta T$ values, together with their discussion in terms of a Random Energy Model, can be found in [67] (see also [69]). Many sets of temperature step experiments of this kind have been performed, by the Saclay group (see references in [45,44]) and also by the Uppsala group (see for example [70]), with similar results, even though sometimes discussed in slightly different terms.

### 5.b Memory dip experiments

The ability of the spin glass to keep a *memory* despite *rejuvenation* has been further explored in experiments with multiple temperature steps. The first "memory dip" experiments, suggested by P. Nordblad, were developed in collaboration between the Uppsala and Saclay groups [71]. An example of a "multiple dip experiment" is shown in Fig.11 [45,72,73].

This is an *ac* experiment in which the sample is cooled by 2 K steps of duration half an hour down to 4 K, and then reheated continuously (sketch in the inset of Fig.11). Fig.11 shows $\chi''$ as a function of temperature during this procedure, starting from $T>T_g$ where $\chi''=0$ (paramagnetic phase). $\chi''$ rises up when crossing $T_g = 16.7\ K$, and when the cooling is stopped at 14 K, the relaxation of $\chi''$ due to aging is recorded during ½ hour (successive points at the same temperature in the figure). Upon further cooling by another 2 K step, a $\chi''$ jump of rejuvenation is found, and the relaxation due to aging takes place. At each new cooling step, rejuvenation and aging are seen, and this happens ~ 6 times in the experiment of Fig.11.

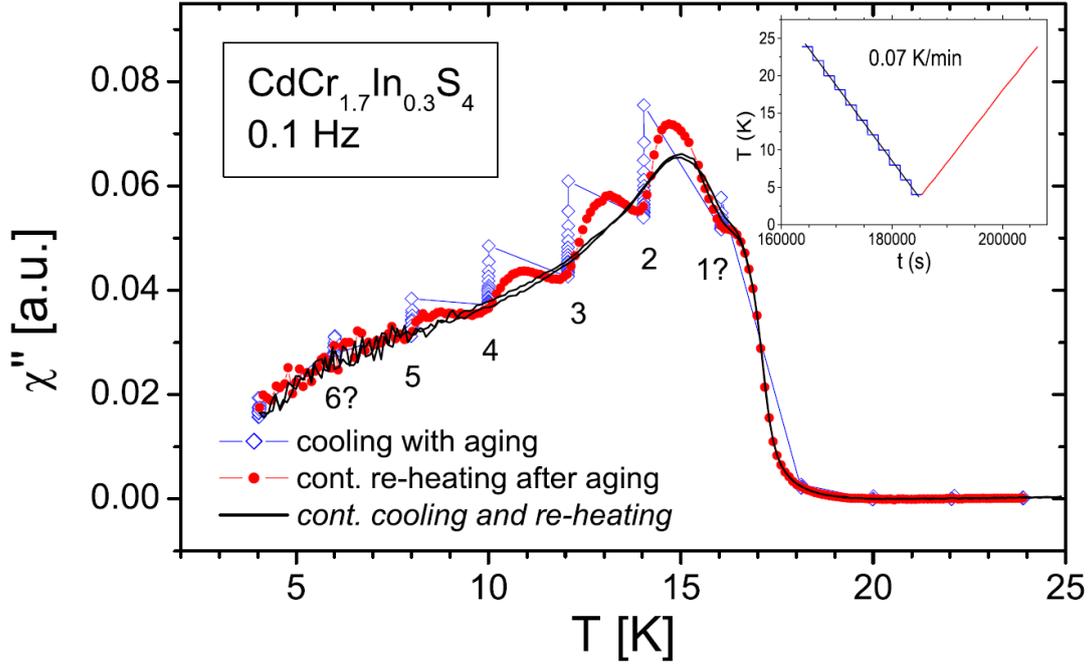

**Figure 11 :** (from [45,72]) An example of multiple rejuvenation and memory steps. The sample was cooled by 2 K steps, with an aging of time of 2000 sec at each step (open blue diamonds). Continuous reheating at 0.001 K/s (full red circles) shows memory dips at each temperature of aging. The solid black line shows a reference measurement with continuous cooling/heating and no steps. The inset is a sketch of the procedures.

In the second part of the experiment, the sample is re-heated continuously at a slow rate (0.001 *K/s*, equal to the average cooling rate). Amazingly, apart from the rather noisy low-*T* region, the memory of each of the aging stages performed during cooling is revealed in shape of "memory dips" in $\chi''(T)$, tracing back the lower value of $\chi''$ which was attained at each of the aging temperatures. Thus, the spin glass is able to keep the simultaneous memory of several (up to 5-6 !) successive aging periods performed at lower and lower temperatures. Increasing the temperature afterwards reveals the memories (and meanwhile erases them).

One can think of a certain type of aging in terms of domain growth dynamics, of the type occurring in a ferromagnet. Aging by domain growth is a naturally "cumulative" process, in the sense that aging continues additively during the various parts of the experimental procedure, from one temperature to the other, as long as $T<T_g$. This cumulative process corresponds to the coherent contribution to aging observed in small temperature step experiments. But it is difficult to imagine how rejuvenation and memory effects may arise in this scheme. In the droplet model [37], they are related to "temperature chaos" effects [74]. Discussions on the possible relevance of this scenario to experiments can be found in [72,75,76], and also [36].

In some spin glass experiments like the one that we present now in Fig.12 [45], the dual aspect of aging dynamics in terms of coherent (cumulative) and incoherent (rejuvenation and memory) contributions appears very clearly. In this *dc* type procedure, proposed by the Uppsala group [77], the sample is cooled in zero-field with various thermal histories, and after applying the field at low temperature the magnetization is measured while increasing the temperature continuously at fixed speed (small steps of 0.1K/min).

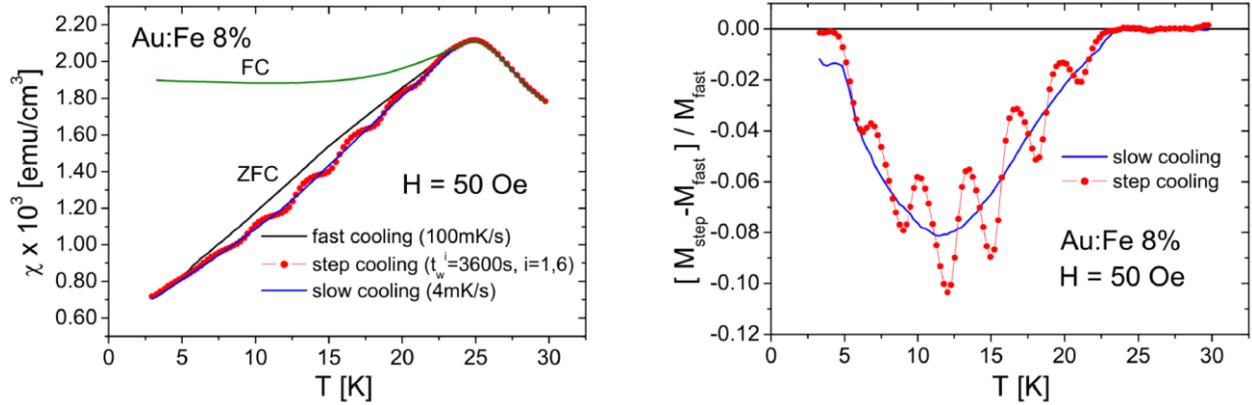

**Figure 12 :** (from [45]) **Left :** Effect of various cooling procedures on the ZFC magnetization of the Au:Fe8% spin-glass. Comparison of fast and slow cooling, with and without stops. **Right :** difference with the magnetization obtained after fast cooling.

On one hand, we can observe the effect of a slow cooling in comparison with that of a fast cooling: the slow-cooled curve lies below the fast one in the whole temperature range. There is indeed a cooling rate effect in spin glasses, provided that one chooses an appropriate procedure to bring it to evidence. On the other hand, memory effects can be demonstrated by stopping the cooling at several distinct temperatures and waiting. The magnetization measured during re-heating after this step-cooling procedure shows clear dips at all temperatures at which the sample has been aging. These effects are emphasized in the right part of Fig.12 where we have plotted the difference between the curves obtained after a specific cooling history and the reference one obtained after a fast cooling. Sharp oscillations (memory dips) show up on top of a wide bump (cumulative aging).

## 5.c Discussion

Thus, aging effects in spin glasses can be described as a combination of rejuvenation and memory effects, which are strongly temperature specific, with some more classical cooling rate effects [56]. Structural glasses are usually considered to be dominated by cooling rate effects [12,13]. However, experiments in various glassy systems have been designed these last years to search for possible rejuvenation and memory effects. Interesting examples of such phenomena can be found for instance in [78,79] with PMMA and in [80,81] with gelatine. New experimental ways have been developed more recently for the investigation of aging effects in colloids and soft matter, like microrheology techniques using optical traps [82,83]. There is now a growing interest for the out-of-equilibrium properties of *active* colloidal systems (chemically powered colloids [84], biomolecular motors [85], Janus particles with asymmetric surface coating [86], etc.)

### 5.c.1 Hierarchical picture

The rejuvenation and memory effects reveal a strong sensitivity of the aging state of the spin glass to temperature changes, which have very different effects when the temperature is decreased or increased. The Saclay group proposed to account for these phenomena in terms of a hierarchical organization of the metastable states as a function of temperature, as pictured in Fig.13, which we now explain ([87,65], see details in [52,44]).

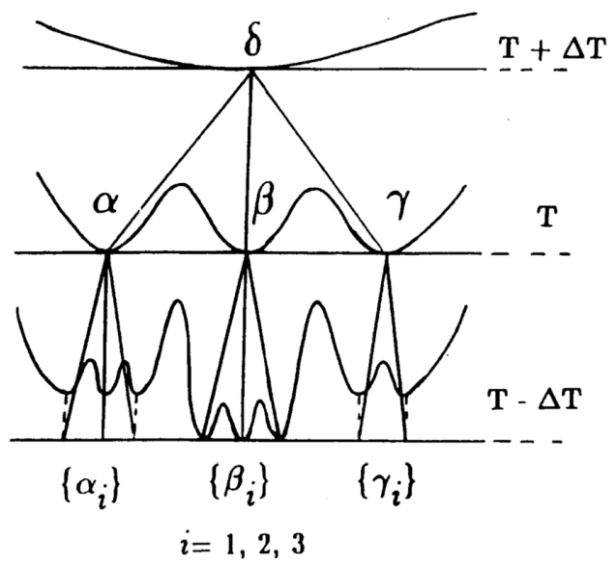

**Figure 13 :** Schematic picture of the hierarchical structure of the metastable states as a function of temperature ([87,65], see details in [52,44]).

In this scheme, the effect of temperature variations is represented by a modification of the free-energy landscape of the metastable states (not only a change in the transition rates between them). At fixed temperature $T$, aging corresponds to the slow exploration of the numerous metastable states (at level $T$ in Fig.13). When going from $T$ to $T$-$\Delta T$, the free-energy valleys subdivide into smaller ones, separated by new barriers (level $T$-$\Delta T$ in Fig.13). *Rejuvenation* arises from the transitions that are now needed to equilibrate the population rates of the new sub-valleys: this is a new aging stage. For large enough $\Delta T$ (and in the limited experimental time window), the transitions can only take place between the sub-valleys inside the main valleys, in such a way that the population rates of the main valleys are untouched, keeping the *memory* of previous aging at $T$. Hence the memory can be retrieved when re-heating and going back to the $T$-landscape.

This tree picture may seem somewhat naïve when described in these qualitative terms. It is, however, able to reproduce many features of the experiments when discussed in more details (see discussions of experimental results in [52,44]). Indeed, quantitative theoretical models have been derived along such a hierarchical scheme, in terms of developments of Bouchaud's Trap Model [88] and also Derrida's Random Energy model [89,67].

The hierarchical organization of the *metastable* states as a function of *temperature* is, of course, reminiscent of the hierarchical organization of the *pure* states (as a function of their *overlap*) that is obtained in the mean-field theory of the spin glass with full replica symmetry breaking [2]. It has been shown that rejuvenation and memory effects can indeed be expected in the dynamics of this model [90]. Detailed analysis of the temperature growth of the free-energy barriers involved in temperature variation experiments has suggested that the hierarchical organization of the *metastable* states as a function of temperature can indeed be extrapolated to a hierarchical organization of the *pure* states [91] (see however the discussion of a different barrier analysis in [72]).

Also, from another point of view [73], rejuvenation effects can be expected from the fact that in a frustrated system effective interactions can be defined, which are found to be temperature dependent in many cases [92]. Still, memory effects require considering other processes in addition [93].

### 5.c.2 A correlation length for spin glass order

Aging can also be considered as the slow establishment of a "spin glass order" [37,38]. Starting from a random configuration obtained when cooling the spin glass from the paramagnetic state (like the structurally liquid state of a glass after quench), the spins will locally optimize their respective orientations over longer and longer length scales, which define a *time growing correlation length*. This correlation length could be determined, although a bit indirectly, in various sets of experiments [94,95].

The rejuvenation and memory effects have important implications in terms of these length scales. Let us consider that during aging at $T$ the correlation length of the spin-glass order grows up to a certain $L_T$. When going to $T-\Delta T$, rejuvenation implies that new reorganization processes take place. But, in order to keep the memory of what happened at $T$, these new processes at $T-\Delta T$ should occur on smaller length scales $L_{T-\Delta T} < L_T$. In practice, the independence of aging at length scales $L_{T-\Delta T}$ and $L_T$ can be realized by a strong separation of the corresponding *time* scales $\tau$:

$$\tau(L, T-\Delta T) \gg \tau(L, T).$$

This necessary separation of the aging length scales with temperature has been coined "temperature-microscope effect" by J.-P. Bouchaud [72]. In experiments like those from Figures 11 and 12, at each temperature stop aging should take place at well-separated length scales

$$L_n < \ldots < L_2 < L_1,$$

as if the magnification of the microscope was varied by orders of magnitude at each temperature step. This hierarchy of embedded length scales as a function of temperature is a *real space* equivalent of the hierarchy of metastable states in the *configuration space* (Fig.13) [72].

Spin glass numerical simulations have allowed the exploration of the microscopic organization of the spins, and investigated the properties of the correlation length of the spin glass state (see for example [96], and references therein). But, due to frustration, the evolutions towards equilibrium are very slow, which implies time consuming computations. The experiments on real spin glasses are typically exploring the $10^0$-$10^5$ s time range, which in units of the paramagnetic spin flip time $\tau_0 \sim 10^{-12}$s corresponds to $10^{12}$-$10^{17}$ $\tau_0$. Taking $\tau_0$=1 MC step for comparison, the first numerical simulations were exploring up to ~$10^7$ Monte-Carlo (MC) steps, a rather short-time regime compared with the experiments. In the Janus and Janus II projects [3], dedicated supercomputers have been designed, which allow computation up to ~$10^{11}$ MC steps. A wide set of numerical results is now available in a time range that is close to that of spin glass experiments, yielding considerable progresses in their interpretation [3].

Intermediate time range of this dynamics could be explored in experiments on the glassy state formed by interacting magnetic nanoparticles (see [97], and references therein). The microscopic flip time of the "super spins" born by the nanoparticles ($10^4$-$10^6$ ferromagnetically coupled spins in each nanoparticle) is much longer than $10^{-12}$ s (depending on the temperature and on the size of the nanoparticles, $\tau_0$ can range from $10^{-10}$ s up to milliseconds), and in $\tau_0$ units the explored experimental time window can be close to that of simulations. The results concerning the time growth of the correlation length $\xi$ of the glassy order during aging in *experiments* (spin and super spin glass) and *simulations* are in overall agreement [94,95,96,97,3]. The general trend of this growth is a slow power law

$$\xi \propto (t/\tau_0)^{aT/T_g},$$

with $a \sim 0.15$, going up to a few tens of atomic distances for spin glasses in the laboratory time window.

## 6. Conclusions

In this paper, we have tried to emphasize some general experimental features of the disordered magnetic systems that are known as spin glasses. The spin glass state develops at and below a well-defined temperature $T_g$, above which the spins form a paramagnetic phase. At $T_g$, the non-linear susceptibility diverges, and the transition to the spin glass state presents most characteristics of a thermodynamic transition. However, since relaxation times are diverging at $T_g$, in an experiment the equilibrium phase cannot be established. Starting from a frozen random configuration inherited from the paramagnetic phase, a "spin glass order" is progressively established at longer and longer range, and this slow evolution is accompanied by aging phenomena that show up in various dynamical properties, like a waiting time dependence of the *ac* susceptibility or of the magnetization relaxation in *dc* field variation procedures.

Lowering the temperature during aging causes a restart of aging processes (*rejuvenation*), while the *memory* of previous aging at higher temperatures can be kept, and retrieved when re-heating. The effect of temperature changes can be seen as a combination of these rejuvenation and memory effects with more common *cumulative* effects. In structural and polymer glasses, and many other glassy systems, the dominant scenario is the continuation of aging from one temperature to another in terms of *cumulative* processes, but rejuvenation and memory processes can often be found, even though with weaker importance. Hence, spin glasses appear as glassy systems in which rejuvenation and memory effects are more pronounced than in others, but they can yet be used as powerful models for glasses in general, because of their rather simple theoretical formulation in terms of a system of randomly interacting spins. Certain classes of spin glass theoretical models (with *p*-spin interactions) have even been found to reproduce rather precisely [98] the properties of structural glasses as modelled by mode coupling theory [99].

Since the early pioneering studies, spin glasses have continuously benefitted from active exchanges between experiments and theory [6,24,30]. They have been the opportunity of important conceptual breakthroughs in statistical physics [2] and these developments have shed new light on the problem of glasses in general [1,13,21]. New results allowing a better microscopic understanding of the transition in structural glasses are now being obtained [18]. The out-of-equilibrium properties of spin glasses have inspired a lot of original experimental investigations of glassy systems in general. After several decades, many crucial questions on the nature of the glassy state are still open, and the important experimental, analytical and numerical efforts that are presently deployed offer very promising perspectives of new progress on these fascinating questions.

# Acknowledgements

The authors are grateful to J.-P. Bouchaud, H. Bouchiat, I. Campbell, H. Kawamura, F. Ladieu, S. Miyashita, S. Nakamae, P. Nordblad, D. Sherrington, Y. Tabata, T. Taniguchi and H. Yoshino for recent discussions and helpful suggestions concerning this manuscript. EV is thankful to the University of Tokyo for kind hospitality during the most part of the writing.